\documentclass[entropy,article,submit,pdftex,moreauthors]{Definitions/mdpi} 
\renewcommand{\linenumbers}{}
\firstpage{1} 
\pubvolume{1}
\issuenum{1}
\usepackage{amsmath}
\DeclareMathOperator\arctanh{arctanh}
\articlenumber{0}
\pubyear{2023}
\copyrightyear{2023}
\datereceived{ } 
\daterevised{ } 
\dateaccepted{ } 
\datepublished{ } 
\hreflink{https://doi.org/} 

\Title{Interplay between Non-Markovianity of Noise and Dynamics in Quantum Systems}

\TitleCitation{Interplay between Non-Markovianity of Noise and Dynamics in Quantum Systems}

\Author{Arzu Kurt $^{*}$\orcidA{}
}

\AuthorNames{Firstname Lastname, Firstname Lastname and Firstname Lastname}

\AuthorCitation{Kurt, A.}

\address{Department of Physics, Bolu Abant \.{I}zzet Baysal University, 14030-Bolu, T\"{u}rkiye
}

\corres{Correspondence: arzukurt@ibu.edu.tr 
}

\abstract{The non-Markovianity of  open quantum system dynamics is often associated with bidirectional interchange of 
information between the system and its environment and is thought to be a resource for various quantum information tasks. We have investigated the 
 non-Markovianity of the dynamics of a two-state system driven by continuous time random walk type noise, 
 which can be Markovian or non-Markovian depending on its residence time distribution parameters. 
 Exact analytical expressions for the distinguishability and the trace-distance and entropy-based 
 non-Markovianity measures are obtained and used to investigate the interplay between the 
 non-Markovianity of the noise and that of dynamics. Our results show that, in many cases, the 
 dynamics are also non-Markovian when the noise is non-Markovian. However, it is possible for Markovian noise to 
 cause non-Markovian dynamics and for non-Markovian noise to cause Markovian dynamics, 
 but only for certain parameter values.}


\begin{document}




\section{Introduction}

Quantum non-Markovianity refers to the existence of memory effects in the dynamics of open quantum systems and has been the subject of many studies with the aim of defining, quantifying, and investigating various schemes to utilize it as a resource for quantum information tasks. Non-Markovianity has been discussed as a possible resource for quantum information tasks such as quantum system control~\cite{Reich2015}, efficient entanglement distribution~\cite{Xiang2014}, perfect state transfer of mixed states~\cite{Laine2014}, quantum channel capacity improvement~\cite{Bylicka2014} and efficiency of work extraction from the Otto cycle~\cite{Thomas2018}. Miller et al.~\cite{Miller2022} have carried out an optical study of the relation between non-Markovianity and the preservation of quantum coherence and correlations, which are essential resources for quantum metrology applications. Various approaches, from environmental engineering to classical driving to controlling the non-Markovianity of quantum dynamics, have been proposed, analyzed, and experimentally realized in recent years. Most non-Markovianity measures invoke bidirectional exchange of information between the system and its environment at the root of the memory effects in the dynamics. The seeming contradiction between such an interpretation and the fact that even external classical noise could induce non-Markovian dynamics~\cite{Pernice2012, Megier2017} was mostly resolved by showing that random mixing of unitary dynamics might lead to memory effects~\cite{Breuer2018, Chen2022}. Representing the quantum environment of a finite-dimensional quantum system using classical stochastic fields has a long history. One of the drawbacks of such an approximation is the effective infinite temperature, which can be resolved by augmenting the master equation with extra terms to restore the correct thermal steady-state. Another seemingly difficult task is to account for the lack of feedback from the system to the classical field. Despite these shortcomings, the stochastic Liouville equation(SLE) approach has produced various interesting physical models of open quantum systems~\cite{Haken1972, Haken1973, Fox1978, Kayanuma1985, Dong2020, Shao1998}.

There have been several studies on the effect of classical noise on the non-Markovianity of quantum dynamics of two-state systems. For example, a study by Cialdi et al. investigated the relationship between different classical noises and the non-Markovianity of the dephasing dynamics of a two-level system~\cite{Cialdi2019}. The study found that non-Markovianity is influenced by the constituents defining the quantum renewal process, such as the time-continuous part of the dynamics, the type of jumps, and the waiting time. In addition, other studies have explored how to measure and control the transition from Markovian to non-Markovian dynamics in open quantum systems, as well as how to evaluate trace- and capacity-based non-Markovianity. It has been shown that classical environments that exhibit time-correlated random fluctuations can lead to non-Markovian quantum dynamics~\cite{Benedetti2014, Benedetti2016}. Costa-Filho et al. investigated the dynamics of a qubit that interacts with a bosonic bath and under injection of classical stochastic colored noise.~\cite{Costa2017} The dynamic decoupling of qubits under Gaussian and RTN was investigated by Bergli et al. \cite{Bergli2007} and~\cite{Cywinski2008}.  Cai et al. have shown that the environment being non-Markovian noise does not guarantee that the system's dynamics are non-Markovian~\cite{Cai2016}.  
When the coupling of the bath to its thermalizing external environment is very strong or on time scales longer than the characteristic microscopic times of the bath, we expect that even fully quantum system-bath models reduce to this case~\cite{Cheng2008}. The addition of non-equilibrium classical noise to dissipative quantum dynamics can be helpful in describing the influence of non-equilibrium environmental degrees of freedom on transport
properties~\cite{Goychuk2004-2}. Goychuk and Hanggi have developed a method to average the dynamics of a two-state system driven by non-Markovian discrete noises of the continuous-time random-walk type (multi-state renewal processes)~\cite{Goychuk2006}.

The transition from Markovian to non-Markovian dynamics via tuning of the system-environmental coupling in various quantum systems has been reported~\cite{Liu2011, Bernardes2014, Brito2015, Garrido2016, Chakraborty2019}.
 The aim of the present study is to provide an answer to the question of whether there is any connection between the non-Markovianity of classical noise and the non-Markovianity of quantum dynamics of a two-state system (TSS) driven by such a noise source. Toward that end, we study the dynamics of a TSS driven by a continuous-time random walk (CTRW) type stochastic process which is characterized by its residence time distribution (RTD) function. We have investigated the effect of biexponential and manifest non-Markovian RTDs. The first one is a simple model of classical non-Markovian noise as a linear combination of two Markovian processes and allows one to study random mixing induced quantum non-Markovianity, while the latter one can be tuned to study a large number of noise models. We have found that exact analytical expressions for the trace-distance and entropic measures of non-Markovianity of the dynamics could be obtained for a restricted set of system parameters. It is well known that Markovian classical noise could lead to non-Markovian quantum dynamics. Here, we have shown that when the driving noise is chosen to be expressively non-Markovian, one can still observe Markovian quantum dynamics depending on the noise and system parameters albeit in a very restricted set. Hence, we have shown that the existence of non-Markovianity in classical noise does not guarantee quantum non-Markovianity of the dynamics of a TSS driven by that noise.

The outline of the paper is as follows. In Section~\ref{model}, we describe the TSS and CTRW noise process and the noise averaging procedure that leads to the exact 
time evolution operator in the Laplace transform domain. The analytical and numerical results of the study for the biased and non-biased TSS for Markovian, as well as the non-Markovian CTRW process, are presented and discussed in Section~\ref{results}. Section~\ref{conc} concludes the article with a brief summary of the main findings.

\section{Model and Non-Markovianity Measures\label{model}}
The main aim of this section is to introduce the TSS model which will be used to study the effect of the non-Markovianity of the classical noise on the non-Markovianity of the quantum dynamics of the TSS driven by the noise and to summarize the trace-distance and entropy-based quantum non-Markovianity measures.
\subsection{Model}
We consider a two-state system (TSS) with Hamiltonian: 
\begin{equation}
H=\frac{1}{2}\hbar \epsilon_0 \sigma_z+\frac{1}{2}\hbar(\Delta_0+\xi(t))\sigma_x+\frac{1}{2}(E_1+E_2)\mathcal{I}
\label{eq:ham}
\end{equation}
\noindent where $\sigma_i$s are the Pauli operators, $E_{1,2}$ are the energies of $|1\rangle$ and $|2\rangle$ states of the TSS, $\Delta_0$ is the static tunneling matrix element, $\epsilon_0=(E_2-E_1)/\hbar$ and $\mathcal{I}$ is the identity operator. The TSS is driven by two-state non-Markovian noise with amplitudes $\xi(t)=\left\{\Delta_{+},\Delta_{-}\right\}$ and stationary-state probabilities 
$p^{st}_{\pm}=\langle\tau_{\pm}\rangle/(\langle\tau_{+}\rangle+\langle\tau_{-}\rangle)$ where $\langle\tau_{\pm}\rangle$ are the average residence time of the noise in states $\Delta_{\pm}$. The stationary autocorrelation function of noise is defined as 
$k(t)=\langle\delta\xi(t)\delta\xi(0)\rangle/\langle[\delta\xi]^2\rangle$ where $\delta\xi(t)=\xi(t)-\langle\xi\rangle_{st}$ and can be expressed in terms of RTDs in the Laplace space as:~\cite{Goychuk2004-2,Goychuk2006}
\begin{eqnarray}
k(s)&=&\frac{1}{s}-\left(\frac{1}{\langle\tau_{+}\rangle}+\frac{1}{\langle\tau_{-}\rangle}\right)\frac{1}{s^2}\frac{\left(1-\psi_{+}(s)\right)\left(1-\psi_{-}(s)\right)}{\left(1-\psi_{-}(s)\psi_{+}(s)\right)}
\end{eqnarray}
\noindent where $\psi_{\pm}(s)$ are Laplace transforms of residence time distribution of the noise in $\Delta_{-}$ and $\Delta_{+}$ states and 
the autocorrelation time of the noise is defined using $k(t)$ as $\tau_{\mathrm{corr}} =\int_{0}^{\infty}|k(t)|\;dt $. If $k(t)$ is strictly positive for all $t$, then $\tau_{\mathrm{corr} } $ can be obtained from $k(s) $ as $\tau_{\mathrm{corr} } =\lim_{s\to 0} k(s) $.

The dynamics of the density matrix $\rho(t)$ of the TSS with the Hamiltonian \ref{eq:ham} can be obtained by expressing it as $\rho(t)=\left[\mathcal{I}+\sum_{i} P_{i}(t)\sigma_{i}\right]/2$ where $P_{i}(t)=\mathrm{Tr}\left[\rho(t)\sigma_i\right] $ is:
\begin{equation}
\dot{P}(t)=F(t) P(t)
\end{equation}
\noindent where $P(t)=\left[P_{x}(t),P_{y}(t),P_{z}(t)\right]^{T}$ and 
\begin{equation}
F[\xi(t)]=\left(\begin{array}{ccc}
-\epsilon_0&0&0\\
\epsilon_0&0&\xi(t)\\
0&\xi(t)&0
\end{array}
\right)
\end{equation}
The noise propagator $S_{\pm}(t)=\exp{\left(F[\Delta_{\pm}]\right)}$ for the static values of noise $\xi=\left\{\Delta_{-},\Delta_{+}\right\}$ as 
\begin{equation}
S_{\pm}(t)=\sum_{k}R_{\pm}^{(k)}\exp{\left(i\lambda_{\pm}^{(k)}t\right)}
\label{eq:prop}
\end{equation}
\noindent where $\lambda_{\pm}^{0}=0 $, $\lambda_{\pm}^{1}=\Omega_{\pm}=\sqrt{\epsilon_0^2+\Delta_{\pm}^2}$ and $\lambda_{\pm}^{2}=-\Omega_{\pm}$ and  
\begin{eqnarray}
R_{\pm}^{(0)}&=&\frac{1}{\Omega_{\pm}^2}\left(
\begin{array}{ccc}
\Delta_{\pm}^2&0&\epsilon_0\Delta_{\pm}\\
0&0&0\\
\epsilon_0\Delta_{\pm}&0&\epsilon_0^2
\end{array}
\right) \nonumber \\
R_{\pm}^{(1)}&=&[R_{\pm}^{(2)}]^{*}=\frac{1}{2}\left(
\begin{array}{ccr}
\frac{\epsilon_0^2}{\Omega_{\pm}^2}&i\frac{\epsilon_0}{\Omega_{\pm}}&-\frac{\epsilon_0 \Delta_{\pm}}{\Omega_{\pm}^2}\\
i\frac{\epsilon_0}{\Omega_{\pm}}&1&i\frac{\Delta_{\pm}}{\Omega_{\pm}}\\
-\frac{\epsilon_0 \Delta_{\pm}}{\Omega_{\pm}^2}&-i\frac{\Delta_{\pm}}{\Omega_{\pm}}&\frac{\Delta_{\pm}^2}{\Omega_{\pm}^2}
\end{array}
\right)
\end{eqnarray}
\noindent 
The problem of obtaining the stationary noise average of the propagator in~(\ref{eq:prop}) involves both averaging over the initial stationary probabilities. It is shown by Goychuk that it can also be done exactly in the Laplace space for non-Markovian processes~\cite{Goychuk2004-2}. The noise-averaged propagator can be expressed as 
\begin{eqnarray}
S(s)&=&p_{+} S_{+}(s)+p_{-} S_{-}(s)-\left(\frac{1}{\tau_{\textcolor{red}{+}}}+\frac{1}{\tau_{\textcolor{red}{-}}}\right)\left\{C_{+}+C_{-}\right.\nonumber\\
&&\left[A_{+}(s)B_{-}(s)+A_{-}(s)\right]\left[I-B_{+}(s)B_{-}(s)\right]^{-1}A_{+}(s) \label{eq:general}\\
&&\left. \left[A_{-}(s)B_{+}(s)+A_{+}(s)\right]\left[I-B_{-}(s)B_{+}(s)\right]^{-1}A_{-}(s)\right\}\nonumber
\end{eqnarray}
\noindent where
\begin{eqnarray}
S_{\pm}(s)&=&\sum_{k}\frac{R_{\pm}^{(k)}}{s-i\lambda_{\pm}^{(k)}} \nonumber\\
A_{\pm}(s)&=&\sum_{k}R_{\pm}^{(k)}\frac{1-\psi_{\pm}\left(s-i\lambda_{\pm}^{(k)}\right)}{s-i\lambda_{\pm}^{(k)}}\nonumber\\
B_{\pm}(s)&=&\sum_{k}R_{\pm}^{(k)}\psi_{\pm}\left(s-i\lambda_{\pm}^{(k)}\right)\nonumber \\
C_{\pm}(s)&=&\sum_{k}R_{\pm}^{(k)}\frac{1-\psi_{\pm}\left(s-i\lambda_{\pm}^{(k)}\right)}{\left(s-i\lambda_{\pm}^{(k)}\right)^2}\nonumber\\
\end{eqnarray}
\noindent where $\psi(s)$ is the Laplace transform of the distribution of the residence time of the noise.
\subsection{Non-Markovianity Measures}
Non-Markovianity of random processes has a well-established and widely accepted definition.
The non-Markovianity of quantum dynamics, on the other hand, although the subject of an immense number of studies in recent years, has not reached a similar consensus. The trace-distance-based measure of non-Markovianity developed by Ref.~\cite{Breuer2009, Breuer2016} quantifies the memory effect in the dynamics with the system's retrieval of information from its environment, which shows up as the nonmonotonic behavior in the distinguishability of quantum states. Given two density operators $\rho_1$ and $\rho_2$, the trace distance (TD) between them is defined as:~\cite{Heinosaari2011}
\begin{equation}
    D(\rho_1,\rho_2)=\mathrm{Tr}\sqrt{(\rho_1-\rho_2)^{\dagger}(\rho_1-\rho_2)}
    \label{eq:td}
\end{equation}
\noindent where $\mathrm{Tr}$ stands for the trace operation. TD is bounded from below as $D(\rho_1,\rho_2)=0$ for $\rho_1=\rho_2$ and from above as $D(\rho_1,\rho_2)=1$ if $\rho_1\perp\rho_2$. As a measure of distinguishability between two quantum states, it can be related to the probability of distinguishing two states with a single measurement~\cite{Fuchs1999}.

Entropy-based Jensen-Shannon divergence (JSD) between two quantum states is another distinguishability measure used to quantify non-Markovianity~\cite{Matjey2005, Settimo2022} and is defined as the smoothed version of relative entropy:
\begin{equation}
    J(\rho_1,\rho_2)=H\left(\frac{\rho_1+\rho_2}{2}\right)-\frac{1}{2}\left(H(\rho_1)+H(\rho_2)\right)
    \label{eq:jsd}
\end{equation}
\noindent where $H(.)$ is the von Neumann entropy $H(\rho)=-\mathrm{Tr} \rho\log{\rho}$. $J(\rho_1,\rho_2)$ has the same bounds as the trace distance in the same limiting cases, but it is not a distance because, contrary to TD, it does not obey the triangle inequality. $\sqrt{J(\rho_1,\rho_2)}$ is shown to be a distance measure~\cite{Virosztek2021} and can be used to quantify the non-Markovianity of quantum dynamics.

Non-Markovianity quantifiers based on a state distinguishability measure $D^{d}(\rho_1,\rho_2)$ are defined as~\cite{Breuer2009, Breuer2016} :
\begin{equation}
    \mathcal{N}^{d}=\max_{\rho_1(0),\rho_2(0)}\int_{\sigma_d(t)>0}\;\sigma_{d}(t)\;dt
    \label{eq:nonmarkov}
\end{equation}
\noindent where
\begin{equation}
    \sigma_{d}(t)=\frac{d}{dt}D^d\left(\rho_1(t),\rho_2(t)\right)
    \label{eq:sigma}
\end{equation}
\noindent where the exponent $d$ stands for either the trace distance distinguishability ($T$) or the Jensen-Shanon entropy divergence ($E$). Maximization in~(\ref{eq:nonmarkov}) is carried out over all possible initial states $\rho_{1,2}(0)$. Wissmann et al.~\cite{Wissmann2012} have shown that $\rho_1(0),\rho_2(0)$ chosen from the antipodal points of the Bloch sphere maximizes the non-Markovianity measure based on the trace distance for two state systems~\cite{Settimo2022, Breuer2009}. For the problem studied, both the trace distance and Jensen-Shannon entropy divergence distinguishability measures could be expressed in terms of population difference $P_z(t)$ and coherences $P_x(t)$ and $P_y(t)$ as 
\begin{eqnarray}
D^{T}&=&\sqrt{P_x^2+P_y^2+P_z^2}\label{eq:dis}\\
D^{E}&=&\frac{1}{\sqrt{\log{4}}}\sqrt{2D^{T}\arctanh{(D^T)}+\log{\left(1-(D^{T})^2\right)}}
\label{std}
\end{eqnarray}

If the chosen distinguishability measure between any two initial states is a monotonic function of time, the dynamics is said to be Markovian. Otherwise, $\mathcal{N}^d$ quantifies the memory effects in dynamics.  

\section{Results and Discussion\label{results}}
We first present the results for TSS whose state energies are degenerate. 
When $\epsilon_0=0$, the Laplace transformed components of the evolution operator can be 
expressed in a simple form as 
\begin{eqnarray}
S_{yy}(s)&=&\frac{s\left(2s^2+\Delta_{-}^2+\Delta_{+}^2\right)}{2\left(s^2+\Delta_{-}^2\right)\left(s^2+\Delta_{+}^2\right)}+\frac{\Delta^2}{\tau}
\left[
\Psi(s)+\Psi^{*}(s)
\right] \label{eq:syy}\\
S_{yz}(s)&=&-\frac{\Delta_0\left(s^2+\Delta_{-}\Delta_{+}\right)}{\left(s^2+\Delta_{-}^2\right)\left(s^2+\Delta_{+}^2\right)}-i\frac{\Delta^2}{\tau}\left[
\Psi(s)-\Psi^{*}(s)
\right]  \\
S_{zz}(s)&=&S_{yy}(s),\,\,\,\,\, S_{zy}(s)=-S_{yz}(s) \label{eq:szz}
\end{eqnarray}
\noindent where 
\begin{equation}
\Psi(s)=\frac{\left[1-\psi\left(s+ i\Delta_{-}\right)\right]\left[1-\psi\left(s+ i\Delta_{+}\right)\right]}{\left(s+ i\Delta_{-}\right)^2\left(s+ i\Delta_{+}\right)^2 \left[1-\psi\left(s+ i\Delta_{-}\right)\psi\left(s+ i\Delta_{+}\right)\right]}
\end{equation}

We will consider a symmetric two-state discrete noise process such that $\Delta_{+}=\Delta=-\Delta_{-}$ is the amplitude, $\tau_{+}=\tau_{-}=\tau$ is the mean residence time and $\psi(s)=\psi_{+}(s)=\psi_{-}(s)$ residence time distribution function of the noise. 
Since one of the aims of the study is to investigate the relation between the non-Markovianity of the driver noise and the quantum dynamics it creates, as the residence time distribution of the noise, we will consider two non-Markovian models, namely bi-exponential and manifest non-Markovian, which have Markovian limiting cases.

\subsection{\label{mark}Markovian noise}
\begin{figure}[!hbt]
    \begin{center}
    \begin{tabular}[b]{c}
    \includegraphics[width=0.4\linewidth]{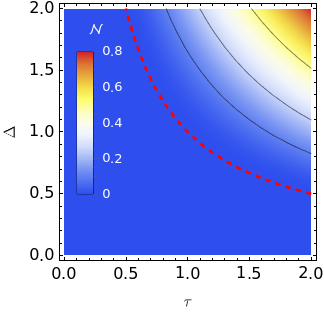}
     \\ \small (a) {Trace distance}
    \end{tabular}\qquad
    \begin{tabular}[b]{c}
    \includegraphics[width=0.4\linewidth]{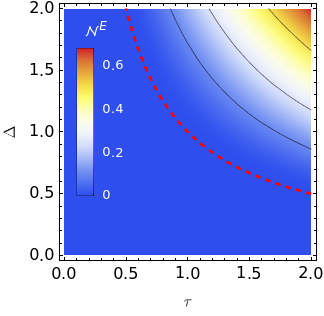}
     \\ \small (a) {Jensen-Shannon divergence}
    \end{tabular}
    \caption{Non-Markovianity of the dynamics for the non-biased TSS as a function of the Markovian noise with auto-correlation time $\tau$ and the amplitude $\Delta$ based on trace distance (a) and Jensen-Shannon divergence distinguishability. The red dotted line is the zero contour while the straight lines denote $\mathcal{N}$ equals (0.1, 0.25, and 0.5).}
    \label{fig:markovian-eps-0}
    \end{center}
\end{figure}

First, we consider the Markovian noise case, which has the RTD $\psi(s)=1/(1+s\tau)$ that can be obtained as $\theta=0,1$ limit of noise with biexponentially distributed residence time (Eq.(\ref{eq:biexp})) or $t_d\to 0$ limit of the manifest non-Markovian RTD (Eq.(\ref{eq:rtd})), both discussed in sections \ref{biexp} and \ref{mani}, respectively. For such an RTD, the inverse Laplace transform of the noise propagators in~(\ref{eq:syy})-(\ref{eq:szz}) can be performed exactly to obtain the following:
\begin{eqnarray}
P_{y}(t)&=&S(t)\;\sin{\left(
\Delta_0 t+\phi
\right)} \label{eq:py}\\
P_{z}(t)&=&S(t)\;\cos{\left(
\Delta_0 t+\phi
\right)} \label{eq:pz}
\end{eqnarray}
\noindent where the initial values of $P_{y}(t)$ and $P_{z}(t)$ are parameterized in terms of $\phi$ as $P_{y}(0)=\sin{\phi}$, $P_{z}(0)=\cos{\phi}$. $S(t)$ in~(\ref{eq:py}) and (\ref{eq:pz}) is the stochastic evolution operator of Markovian two-state noise:
\begin{equation}
S(t)=e^{-t/\tau}\;\left[\cosh{\left(\sqrt{1-\Delta^2\tau^2}\;t\right)}+\frac{1}{\sqrt{1-\Delta^2\tau^2}}\sinh{\left(\sqrt{1-\Delta^2\tau^2}\;t\right)}\right]
\end{equation}
The trace distance distinguishability of the dynamics can be calculated from~(\ref{eq:dis}) by inserting population and coherence expressions from~(\ref{eq:py}) and (\ref{eq:pz}) as follows:
\begin{equation}
D(\rho_1,\rho_2)=\left|S(t)\right|
\label{eq:dist}
\end{equation}
\noindent One should note that $S(t)$ is a monotonously decreasing function of $t$ for  $\Delta \tau<1$ but displays decaying oscillations when $\Delta \tau>1$ as hyperbolic trigonometric functions inside the parentheses transform to ordinary trigonometric functions when $\Delta \tau>1$. Since the non-Markovianity measure (Eqs.~\ref{eq:nonmarkov} and \ref{eq:sigma}) is defined as the integral of positive values of the time derivative of $\mathcal{D}$, $\mathcal{N}=0$ for $\Delta\,\tau<1$. Interestingly, the trace distance distinguishability-based non-Markovianity measure for this particular $\mathcal{D}$ and $\Delta\,\tau>1$ can be obtained analytically in a simple form as 
\begin{equation}
\mathcal{N}=\frac{1}{e^{\frac{\pi}{\sqrt{\Delta^2\tau^2-1}}}-1}
\label{eq:nm}
\end{equation}
\noindent Here, the non-Markovianity is found to be independent of the static value of the coupling coefficient $\Delta_0$. A similar expression for $\mathcal{N}$ has been reported in  Ref.~\cite{Benedetti2014}  for a similar Markovian two-state noise. It is also easy to obtain an analytical expression for the Jensen-Shannon entropy divergence for the present case as follows:
\begin{equation}
J(t)=\frac{1}{\log{4}}\left\{\log{\left[1-S^2(t)\right]} +2S(t)\arctanh{[S(t)]}\right\}
\label{eq:markjon}
\end{equation}
\noindent Although it is possible to derive an exact expression for entropy-based non-Markovianity measure by using~(\ref{eq:nonmarkov}) and (\ref{eq:markjon}), the expression is not compact enough to be helpful in deciphering the relation between $\mathcal{N}^{E}$ and the noise parameters. Therefore, we display only the calculated 
entropy-based $\mathcal{N}^{E}$ along with the one derived from the trace distance distinguishability in Figure~\ref{fig:markovian-eps-0}.
The contours of non-Markovianity are plotted in Figure~\ref{fig:markovian-eps-0} as functions of the mean residence time $\tau$ and noise amplitude $\Delta$. As can be seen from~(\ref{eq:nm}) and the plot, $\mathcal{N}$ is nonzero as long as the Kubo number of the noise is greater than one, which is known as slow noise or strong system-noise coupling or strongly colored noise regime~\cite{Zhou2010}.
Interestingly, both measures are found to signal the same limits ($\Delta\tau>1$) for the existence of non-Markovianity in the dynamics. Furthermore, even the magnitudes of $\mathcal{N}$ and $\mathcal{N}^{E}$ are found to be comparable. We have observed the same behavior for all the other noise models reported in the following, and for the remainder of the paper, we will report results only for the trace-distance-based measure $\mathcal{N}$. 

An interesting dynamics and non-Markovianity behavior is observed if the noise RTD is chosen as the $\alpha\to 0$ limit of the manifest non-Markovian RTD in~(\ref{eq:rtd}) which reduces $\psi(s)$ to a form similar to that of Markovian noise with a modified mean residence time. It is easy to perform an exact analytical inverse Laplace transform of the propagator expressions in~(\ref{eq:syy})-~(\ref{eq:szz}) for $\psi(s)=1/(1+s\tau\tanh(1))$ and find the population difference as:
\begin{equation}
P_z(t)=\frac{1}{1+e^2}\left(2 \cos (\Delta  t)+
\left(e^2-1\right) S_2(t)\right)
\end{equation}
where
\begin{equation}
S_2(t)=e^{-ct/\tau}\left(\cosh{\left(t C/\tau\right)}+\frac{1+e^2}{\sqrt{\left(1+e^2\right)^2-(e^2-1)^2 \Delta^2\tau^2}}\sinh{\left(t C/\tau\right)}\right)
\end{equation}
\noindent where $C=\sqrt{\coth^2{1}-\Delta^2\tau^2}$ and $c=\coth{1}$. As $t$ approaches infinity, $S_2(t)$ approaches zero, while $P_z(t)$ exhibits oscillations with an amplitude of $2/(1+e^2)$ and a frequency $\Delta$. The non-Markovianity of the dynamics, as assessed by both the trace distance and Jensen-Shannon entropy, is found to be unbounded. It is worth noting that the long-term limit of $P_z(t)$ is insensitive to both the noise amplitude $\Delta$ and the mean residence time $\tau$. This result contradicts the findings obtained for Markovian noise for which we have found that $\mathcal{N}$ is zero for $\Delta\tau<1$ and tends to a finite value for $\Delta\tau>1$. It should be noted that $\alpha\to 0$ limit of manifest non-Markovian process describes a noise with $1/\omega$ power spectrum~\cite{Goychuk2004} near $\omega=0$, which is similar to widely studied $1/f$ noise. Benedetti et al. studied~\cite{Benedetti2014} the non-Markovianity of colored $1/f^{\alpha}$ noise-driven quantum systems and reported finite values for $\mathcal{N}$ in contrast to our findings.

\subsection{\label{biexp}Biexponentially distributed residence time}
Biexponential RTD in the time domain is defined as:~\cite{Goychuk2004} 
\begin{equation}
    \psi(t)=\theta \alpha_1 \exp{\left(-\alpha_1 t\right)}+(1-\theta) \alpha_2 \exp{\left(-\alpha_2 t\right)}
    \label{eq:biexp} 
\end{equation}
where $\theta$ and $(1-\theta)$ are the probabilities of the realization of the transition 
rates $\alpha_1$ and $\alpha_2$. The mean residence and autocorrelation times of this noise can be expressed as
\begin{eqnarray}
    \langle\tau\rangle&=&\theta/\alpha_1+(1-\theta)/\alpha_2\\
   \tau_{\mathrm{corr}}&=&\int_0^{\infty}|k(t)|\;dt
\end{eqnarray}
\noindent $\theta=0$ and $\theta=1$ correspond to Markovian noise with mean residence times $1/\alpha_1$ and $1/\alpha_2$, respectively. 
The two-state noise with biexponential residence time distribution allows one to define a non-Markovianity quantifier, denoted by $C_V$, which can be tailored by tuning the parameter $\theta$. This quantifier is given by the ratio of the mean autocorrelation time of non-Markovian noise, $\langle \tau_{\mathrm{corr}}\rangle=\int_{0}^{\infty}k(t)dt$, to the autocorrelation time of the Markovian process $\tau_{\mathrm{corr}}^{M}=\langle\tau\rangle/2$ through the mean residence time $\langle\tau\rangle$ as in~(\ref{eq:nonm}):
\begin{equation}
C_{V}^2=\frac{2}{\langle\tau\rangle}\tau_{corr}
\label{eq:nonm}
\end{equation}

The Laplace transformed expressions for the noise propagator in~(\ref{eq:syy})-~(\ref{eq:szz}) for the biexponential RTD are amenable to be transformed back to the time domain for the nonbiased TSS. But the resulting population, coherence, and trace distance expressions are tedious to display here. On the other hand, for the manifest non-Markovian RTD, the only way to perform the inverse transformation is to use numerically exact inverse Laplace transformation (ILT) methods. We have tested CME~\cite{Horvath2019}, Crump~\cite{Crump1976}, Durbin~\cite{Durbin1973}, Papoulis~\cite{Papolis1957}, Piessens~\cite{Piessens1975}, Stehfest~\cite{Stehfest1970}, Talbot~\cite{Talbot1970}, and Weeks numerical ILT algorithms and have found that the method based on concentrated matrix exponential (CME) distributions reported in~\cite{Horvath2019} has the best performance in terms of computational cost for a given accuracy. The convergence of the computed quantities as a function of the number of included terms and the working precision is carefully checked, and 300 terms and 64-bit precision are found to be adequate for all the reported calculations to converge to 0.1\%.

\begin{figure}[!htb]
\begin{center}
    \begin{tabular}[b]{c}
\includegraphics[width=0.45\linewidth]{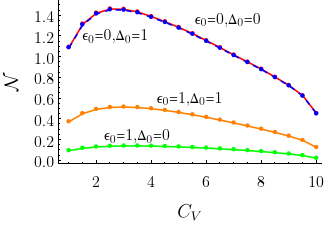}
\\ \small (a) {Non-Markovianity}
    \end{tabular}
    \begin{tabular}[b]{c}
    \includegraphics[width=0.45\linewidth]{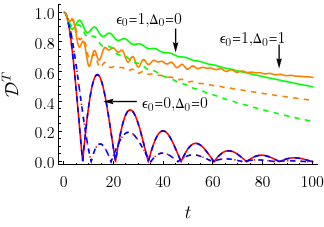}
    \\ \small (b) {Trace distance}
    \end{tabular}   
    \caption{Noise non-Markovianity $C_V$ dependence of the trace-distance based non-Markovianity measure $\mathcal{N}$ (a) and trace-distance distinguishability $\mathcal{D}^T$ (b) for the two-state discrete noise with bi-exponential residence time distribution. The noise parameters are $\Delta=1/4$, $\alpha_1=1/20$, and $\alpha_2=1$. $\theta$s are chosen such that $C_V$ ranges from 1 to 10. $\mathcal{N}$ and $\mathcal{D}^T$ for four combinations of TSS transition energy $\epsilon_0$ and electronic coupling $\Delta_0$ values are displayed. Note that for the nonbiased case ($\epsilon_0=0$), the difference in $\mathcal{N}$ between $\Delta_0=0$ and $\Delta_0=1$ is minimal and indistinguishable on the plots. The straight (dashed) lines in $\mathcal{D}^T$ plots of (b) are calculated at $C_V=4$ (10).}
\label{fig:biexp}
\end{center}
\end{figure}
$\mathcal{N}$ of TSS dynamics as a function of noise non-Markovianity parameters $C_{V}$ is shown in Figure~\ref{fig:biexp}a for noise amplitude $\Delta=1/4$ with $\Delta_0=0,1$ and $\epsilon_0=0,1$.  Remarkably, it is observed that for the four combinations of the site energy difference $\epsilon_0$ and the static coupling $\Delta_0$, the non-Markovianity of quantum dynamics displays a broad resonance structure as a function of $C_{V}$ that indicates that increasing the non-Markovianity of the classical driving noise beyond a certain threshold would decrease the non-Markovianity of the driven quantum dynamics. Figure~\ref{fig:biexp}b shows the trace-distance distinguishability at two chosen $C_{V}$ values and indicates that the main effect of increasing $C_{V}$ is to increase the dissipation rate of the dynamics. These results indicate that the increasing non-Markovian nature of the driving noise might increase, but also decrease the non-Markovianity of the quantum dynamics of the system studied depending on the magnitude.

\subsection{\label{mani}The manifest non-Markovian noise}
The other residence time distribution, we will investigate, is a manifest non-Markovian noise with RTD defined in the Laplace space as:~\cite{Goychuk2004, Goychuk2006} 
\begin{equation}
\psi(s)=\frac{1}{1+s\tau g(s)} 
\label{eq:rtd}
\end{equation}
\noindent with
\begin{equation}
g(s)=\frac{\tanh{\left[\left(s t_{d}\right)^{\alpha/2}\right]}}{\left(s t_{d}\right)^{\alpha/2}}
\end{equation}
\noindent $\tau$ is the mean residence time of the noise and $t_{d}$ is another time constant that can be used to control the non-Markovianity of the noise (at the limit $t_{d}$ = 0, $\psi(t)$ is exponential). The parameter $\alpha$ which is limited to the range $0<\alpha<1$ characterizes the noise-power distribution: $\psi(s)$ describes noise that shows $1/\omega^{1-\alpha}$ features in its spectrum as $\omega\to 0$ and encompasses various power-law residence time distributions. $\alpha=1$ describes normal diffusion, while the $0<\alpha<1$ case corresponds to subdiffusion with index $\alpha$ in the transport context~\cite{Goychuk2004}. 
One of the interesting properties of discrete, manifestly non-Markovian noise is that its correlation time is infinite for $\alpha<1$, which means that the Kubo number is effectively infinite, and no perturbative treatment would produce any reasonable accurate dynamics. The current method based on the Laplace transform is the only way to investigate the dynamics for such residence-time distributions. We have discussed the two limiting cases, namely $t_d\to 0$ (Markovian) and $\alpha\to0$ (infinite $\mathcal{C}$), of the manifest non-Markovian RTD above. Here, we present and discuss how the RTD parameters $\alpha$ and $t_d$ affect the trace distance distinguishability and non-Markovianity of the TSS dynamics at different system parameters. 

\begin{figure}[!htb]
\begin{center}
    \begin{tabular}[b]{c}
    \includegraphics[width=0.32\linewidth]{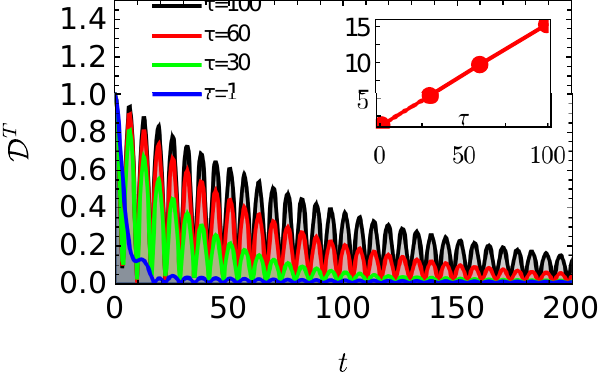}
   \\ \small (a) {$\epsilon_0=0$, $t_d=1$}
    \end{tabular}
    \begin{tabular}[b]{c}
    \includegraphics[width=0.32\linewidth]{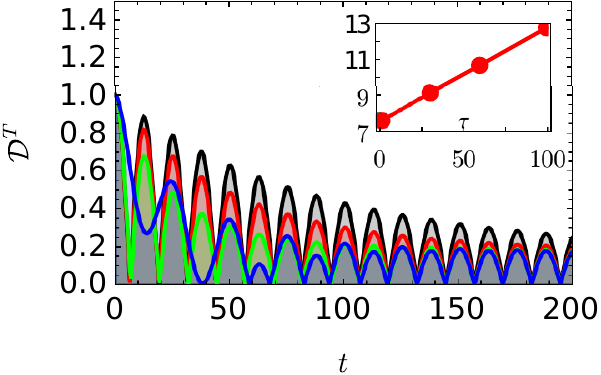}
   \\ \small (b) {$\epsilon_0=0$, $t_d=10$}
    \end{tabular}
     \begin{tabular}[b]{c}
    \includegraphics[width=0.32\linewidth]{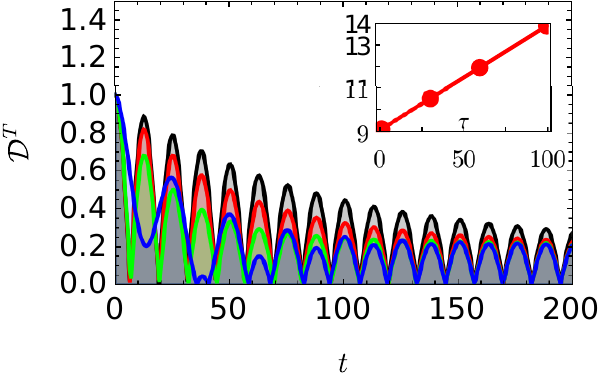}
   \\ \small (c) {$\epsilon_0=0$, $t_d=100$}
    \end{tabular}\qquad
     \begin{tabular}[b]{c}
    \includegraphics[width=0.32\linewidth]{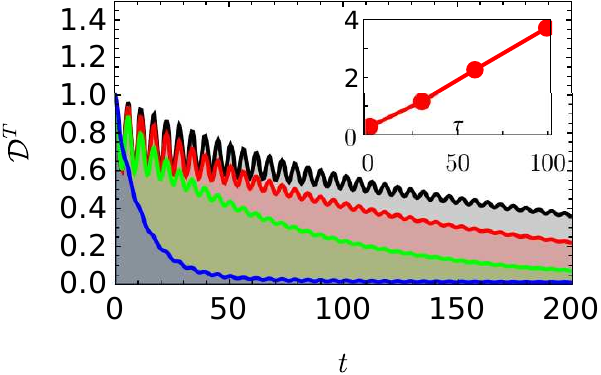}
   \\ \small (d) {$\epsilon_0=1$, $t_d=1$}
    \end{tabular}
    \begin{tabular}[b]{c}
    \includegraphics[width=0.32\linewidth]{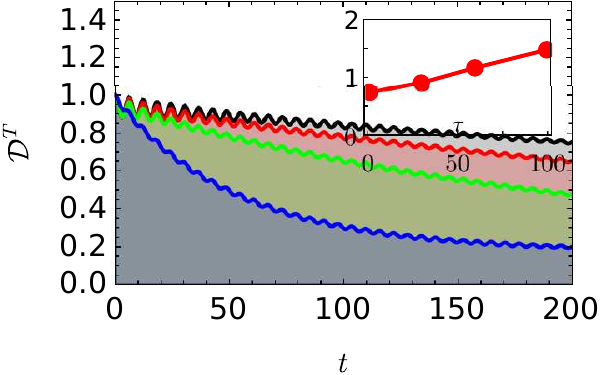}
   \\ \small (e) {$\epsilon_0=1$, $t_d=10$}
    \end{tabular}
    \begin{tabular}[b]{c}
    \includegraphics[width=0.32\linewidth]{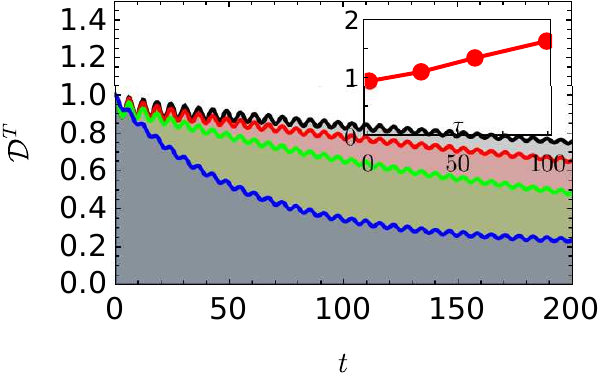}
   \\ \small (f) {$\epsilon_0=1$, $t_d=100$}
    \end{tabular}
    \caption{Trace-distance as a function of time for the manifestly non-Markovian noise at different $t_d$ parameters and average residence time $\tau$. Insets show the trace distance-based non-Markovianity measure as a function of $\tau$. The other parameters of the noise and the system are $\alpha=1/2$, $\Delta_0=0$, and $\Delta=1/2$.}
    \label{fig:nm-td1}
    \end{center}
\end{figure}

First, we present the trace distance distinguishability along with the associated non-Markovianity $\mathcal{N}$ for the manifestly non-Markovian noise for various $t_d$ and mean residence time $\tau$ in Figure~\ref{fig:nm-td1} for biased and nonbiased TSS at $\alpha=0.5$ and $\Delta=0.5$. As $t_{d}$ is a rough measure of the non-Markovianity of manifest non-Markovian noise, one can infer, from a comparison of insets in Figures~\ref{fig:nm-td1}a and ~\ref{fig:nm-td1}c as well as Figures~\ref{fig:nm-td1}c and Figure~\ref{fig:nm-td1}d, that $\mathcal{N}$ increases with increasing $t_d$ for both nonbiased and biased TSS. The mean residence time dependence of $\mathcal{N}$ is found to be independent of  $t_{d}$. $\mathcal{N}$ increases with increasing $\tau$ for all three values considered in this work for the biased as well as the non-biased TSS. Furthermore, $\mathcal{N}$ of the biased case is always found to be lower than that of the nonbiased case. Another interesting observation from Figure~\ref{fig:nm-td1}b is that the trace-distance distinguishability for TSS driven by the highly non-Markovian noise tends to a non-zero constant instead of the expected zero.  

\begin{figure}[!htb]
\begin{center}
    \begin{tabular}[b]{c}
    \includegraphics[width=0.45\linewidth]{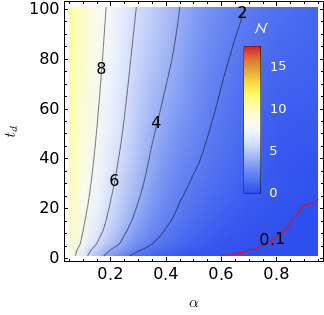}
   \\ \small (a) {$\Delta=0.1,\,\tau=1$}
    \end{tabular}
    \begin{tabular}[b]{c}
    \includegraphics[width=0.45\linewidth]{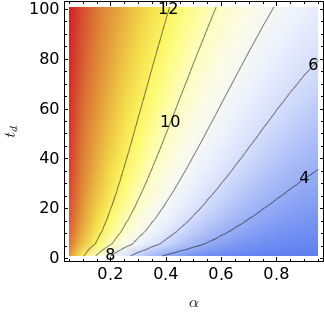}
   \\ \small (b) {$\Delta=0.5,\,\tau=20$}
    \end{tabular}
     \caption{$\alpha$ and $t_{d}$ dependence of trace-distance based non-Markovianity $\mathcal{N}$ of the dynamics of TSS driven with manifest non-Markovian two state noise at different Kubo numbers $K=0.1$ (a) and $K=10$ (b). The same color map is used for both plots and the iso-$\mathcal{N}$ values are shown as the contour labels. The red contour line in (a) is the $\mathcal{N}=0.1$ contour. }
    \label{fig:nm-td-alpha}
    \end{center}
\end{figure}

To further delineate the relationship between $\mathcal{N}$ and the noise parameters $\alpha$ and $t_{d}$, we present the trace-distance-based non-Markovianity measure $\mathcal{N}$ as a function of the exponent $\alpha$ and the $t_{d}$ time parameter of the noise residence time distribution for the dynamics of nonbiased TSS in Figure~\ref{fig:nm-td-alpha} in two different combinations of noise amplitude and mean residence time. The mean residence time of the noise is $\tau=1,\,20$ in these graphs, and the amplitude of the noise is chosen as $\Delta=0.1,\,0.5$ for the subgraphs. The most important observation from Figure~\ref{fig:nm-td-alpha} is that the Kubo number is the most important noise parameter that determines the magnitude of the non-Markovianity of the TSS dynamics. The larger $\Delta$ leads to a larger $\mathcal{N}$ for given $\alpha$ and $t_{d}$ values. This finding is similar to the one we have discussed above for Markovian noise; the existence of non-Markovianity, in that case, depends on if $\Delta\tau>1$. For the manifest non-Markovian noise, the dynamics are found to be non-Markovian even for $\Delta \tau<1$. But the magnitude of $\mathcal{N}$ still strongly depends on the Kubo number $K=\Delta\tau$. Figure~\ref{fig:nm-td-alpha} also indicates that $\mathcal{N}$ depends on $t_{d}$ weakly above a threshold (around $t_{d}=15$) and $\mathcal{N}$ increases smoothly with $\alpha$ for constant $t_{d}$ in most of the $\alpha-t_{d}$ plane. It should also be noted that $\mathcal{N}$ can be zero under manifest non-Markovian noise driving when $\alpha\to 1$ when $\Delta\ll 1$. This limit corresponds to white noise with a constant power spectrum at all frequencies.  

\section{Conclusion\label{conc}}

We have studied Jensen-Shannon entropy divergence and trace distance-based measures of non-Markovianity of dynamics of a two-level system under continuous-time random walk-type stochastic processes with Markovian and non-Markovian residence-time distributions to delineate whether there is any connection between Markovianity of the noise and that of dynamics. We were able to obtain analytically exact expressions for both measures for the nonbiased TSS driven by Markovian CTRW noise. This expression indicates that, above a critical Kubo number of the noise, even Markovian noise can lead to non-Markovian quantum dynamics. The numerical study of biased TSS with the same external noise is found to be mainly smearing of the exact boundary between the Markovian-non-Markovian boundary in the noise frequency-noise amplitude or the classical noise-TSS coupling coefficient plane. We have used non-Markovian noise with biexponential distribution as a model of non-Markovianity produced by random mixing of Markovian dynamics and found that increasing the non-Markovianity of the noise might not lead to increased $\mathcal{N}$ for the dynamics. We have also considered a CTRW with manifest non-Markovian residence-time distribution and shown that the dynamics can be Markovian even for such a noise. An interesting finding of the study was obtained at the $\alpha\to 0$ limit of manifest non-Markovian noise. The exact expression obtained for the trace distance at this limit showed that $\mathcal{N}$ is infinite at this limit. As the discussion on the proper definition and measure of non-Markovianity of quantum dynamics has not been settled yet, the results reported in this study provide a case study for answering the "does the non-Markovianity of the classical driver determine the non-Markovianity of the driven?" question.  





\funding{This study was supported by the Scientific and Technological Research Council of T\"{u}rkiye (TUBITAK) Project no. 1002-120F011.}

\dataavailability{Data are available from the author upon reasonable request.} 

\acknowledgments{The author acknowledges many useful comments and discussions with Prof. Dr. Resul Eryi\u{g}it.}

\conflictsofinterest{The author declares no conflict of interest.} 



\reftitle{References}

\end{document}